\documentclass[%
  aps,
  prl,
  superscriptaddress, letterpaper,
  twocolumn,
  amsfonts, amssymb, amsmath]{revtex4-1}

\usepackage{xcolor}
\usepackage{inputenc}
\usepackage{graphicx,amssymb,amsmath,array,multirow,soul}
\usepackage{hyperref} 
\hypersetup{pdfpagemode={UseOutlines},
bookmarksopen=true,
bookmarksopenlevel=0,
hypertexnames=false,
colorlinks=true, 
citecolor=blue, 
linkcolor=red, 
urlcolor=black, 
pdfstartview={FitV},
unicode,
breaklinks=true,
}

\definecolor{pink}{rgb}{1,1,0} 
\definecolor{red}{rgb}{1,0,0}
\definecolor{yellow}{rgb}{1,1,0}
\definecolor{orange}{rgb}{1,0.5,0} 
\definecolor{blue}{rgb}{0,0,1}
\definecolor{white}{rgb}{1,1,1}

\def\NOTE#1{{\textcolor{red}{\bf [#1]}}}   
\def\NOTE2#1{{\textcolor{blue}{\bf [#1]}}}   

\begin{document}

\title{Experimental evidence of a hydrodynamic soliton gas 
}
\author{Ivan Redor}
\author{Eric Barth\'elemy}
\author{Herv\'e Michallet}
\affiliation{Laboratoire des Ecoulements Geophysiques et Industriels, Universite Grenoble Alpes, CNRS, Grenoble-INP,  F-38000 Grenoble, France}
\author{Miguel Onorato}
\affiliation{Dipartimento di Fisica, Universit\`a di Torino and INFN, 10125 Torino, Italy}
\author{Nicolas Mordant}
\email[]{nicolas.mordant@univ-grenoble-alpes.fr}
\affiliation{Laboratoire des Ecoulements Geophysiques et Industriels, Universite Grenoble Alpes, CNRS, Grenoble-INP,  F-38000 Grenoble, France}

\begin{abstract}
We report on an experimental realization of a bi-directional soliton gas in a 34~m-long wave flume 
in shallow water regime.   We take advantage of the fission of a sinusoidal wave to inject continuously  solitons that propagate along the tank, back and forth. 
Despite the unavoidable damping,  solitons retain adiabatically their profile, while decaying. The outcome is the formation of a stationary state characterized 
by a dense soliton gas whose statistical properties are well described by a pure integrable dynamics. The basic ingredient in the gas, i.e. the two-soliton interaction, 
is studied in details and compared favourably with the analytical solutions of the Kaup-Boussinesq integrable equation. High resolution space-time measurements 
of the surface elevation in the wave flume provide a unique tool for studying experimentally the whole spectrum of excitations.
\end{abstract}


\maketitle
In $1965$ Zabusky and Kruskal coined the word ``soliton" to characterize  two pulses  that 
``shortly after the interaction, they reappear virtually unaffected in size or
shape'' \cite{zabusky1965interaction}.
This property, that makes solitons fascinating objects, is a common feature of solutions of  integrable equations, such as for example the celebrated 
Korteweg-de Vries (KdV) equation that describes long waves in dispersive media,  or the Nonlinear Shr\"odinger  equation, suitable for describing cubic nonlinear, narrow-band processes. Those equations find applications in many fields of physics such as nonlinear optics, water waves, plasma waves, condensed matter, etc \cite{Dauxois}. 
In analogy to a gas of interacting particles described mesoscopically by the classical Boltzmann equation, 
in the presence of a large number of interacting solitons, Zakharov in 1971 derived a  kinetic equation 
for the velocity distribution function of solitons  \cite{zakharov1971kinetic}, see also \cite{el2005kinetic,el2011kinetic,El}. 
Some of the theoretical predictions have been confirmed {\it via} numerical simulations of the KdV equation  in \cite{carbone}.  The wave-counterpart of the particle-like interpretation of solitons is known as  ``integrable wave turbulence''; such a concept was introduced more recently by Zakharov \cite{Zakharov}. The major question in this field is again the understanding of the statistical properties of an interacting ensemble of nonlinear waves, described by integrable equations, in the presence or not of randomness; the latter may arise from initial conditions which evolve under the coaction of   linear and nonlinear effects, \cite{Agafontsev,walczak2015optical,agafontsev2016integrable,soto2016integrable,randoux2016nonlinear,randoux2014intermittency,randoux2017optical,Pelinovsky,schwache1997properties}. In contrast to many nonintegrable closed wave systems that reach a thermalized state characterized by the equipartition of energy among the degrees of freedom (Fourier modes) \cite{onorato2015route,pistone2018thermalization}, integrable equations are characterized by an infinite number of conserved quantities and their dynamics is confined on special surfaces in the phase space. This prevents the phenomenon of classical thermalization and it opens up the fundamental quest on what is the large time state of integrable systems for a given class of initial conditions. So far the question has no answer and, apart from recent theoretical approaches \cite{dyachenko2016primitive,girotti2018rigorous}, most of the results on the KdV problem relies on numerical simulations, see e.g. \cite{Pelinovsky,carbone,Dutykh:2014dm,Pelinovsky:2017ce}. 
{
Soliton gas have been extensively observed in optics \cite[e.g.][]{mussot2009observation,dudley2014instabilities}  while experimental evidence in a hydrodynamic context is scarce.
}
In \cite{Costa} it has been claimed that the low frequency component of sea surface elevations measured in the Currituck Sound (NC, USA) behave as a dense soliton gas, displaying a power law energy spectrum with exponent equal to -1; 
another approach is described in \cite{giovanangeli} where the soliton content in laboratory shallow water wind waves is estimated.

In this Letter we describe a unique experiment that is designed to build and monitor a hydrodynamic soliton gas in a laboratory. 
We focus on shallow water gravity waves where the dynamics is described to the leading order in nonlinearity and dispersion by the  KdV equation. 
In planning the experiment, many issues had to be faced. The main one is that dissipation is present in any experimental set-up; 
thus, the integrable equations cannot describe experiments over long time scales. In order to reach a stationary regime, energy must be injected by a forcing device. 
These two features, dissipation and forcing, are clearly at odds with the integrable turbulence framework. 
Therefore, is it possible to produce in the laboratory  a soliton gas described in statistical terms by  integrable turbulence? 
Answering positively this question in laboratory experiments would strongly support the application of the integrable turbulence framework to in-situ data, 
as reported in~\cite{Costa,Osborne}. 

 Experiments are performed in a $34\,$m-long wave flume, $55\,$cm-wide with a water depth at rest equal to $h=12\,$cm.
 Waves are generated by a piston-type wavemaker (see a more detailed description in~\cite{Guizien,Guizien2}). 
 The free surface vertical displacement is measured along the flume by imaging through the lateral glass walls. 
 We use seven synchronized monochrome cameras with resolution $1920\times1080$ pixels; each camera records the image of the contact line of water over 2~m. 
 Thus, the field of view consists of the 14~m-long central part of the flume. The spatial resolution is close to 1~mm/pix. 
 A grid is used to calibrate the images and correct their geometrical distortions. 
 The contact line is extracted by detecting the strongest grayscale gradient in a vertical line of pixels. 
 Subpixel resolution is obtained by polynomial interpolation in the vicinity of the pixel of steepest gradient. 
 We estimate that the resulting error on the surface elevation is close to $0.1\,$mm. The cameras are operated at $20\,$frames/s. 

The second important issue to deal with during  the design of a soliton gas experiment is  that wave flumes  
are usually not long enough to observe many soliton collisions in the KdV regime which account only for wave propagating in one direction. 
To cope with such limitation, the end of the flume, opposite to the piston,  consists of a vertical wall that reflects the waves. 
Thus the waves propagate in both directions and are also reflected on the piston. 
We take advantage of the reflections to observe the propagation over time scales larger than the duration restricted to a one-way trip along the flume. 
An integrable system of equations that deals with bidirectional wave propagation in shallow water is the Kaup-Boussinesq (KB) system:
\begin{eqnarray}
\partial_t\eta+\partial_x\left[(h+\eta)u\right]&=&-\frac{h^3}{3}\partial_{xxx}u \nonumber\\
\partial_tu+u\partial_xu+g\partial_x\eta&=&0,
\label{KB}
\end{eqnarray}
where $\eta=\eta(x,t)$ is the free surface  elevation, $u=u(x,t)$ is the fluid velocity, $g$ is the acceleration of gravity and 
$h$ is the water depth. Single soliton, as well as multi-solitons solutions, are known for this system and reported in~\cite{Zhang}. 
The multi-solitons solutions of KB include the collision of two solitons: either overtaking collisions 
(two solitons of distinct amplitude propagating in the same direction) or head-on collisions. 
Such collisions are elastic, i.e. the amplitudes of the solitons are not altered after the interaction. The effect of the collision is to induce a phase shift. 
For head-on collisions such a phase shift is much smaller than for overtaking solitons
For the latter, the larger solitons ``jump'' forward, while the smaller solitons experiences a negative delay. 
For co-propagating solitons, the collision is very close to the KdV one \cite{Zhang}.

{The amplitude of a single soliton traveling in the flume 
decays exponentially with an e-fold time scale of $90\,$s (see supplemental material)}, i.e. a clear indication of the presence of viscous dissipation 
(in the bulk and along the walls and bottom) and of the non integrable long term dynamics. 
Nevertheless, the shape of the soliton is fitted correctly by the single soliton solution, 
suggesting that the shape of the soliton changes adiabatically and highlighting the short time integrable dynamics~\cite{keulegan1948}.

\begin{figure}
\includegraphics[viewport=80 210 525 660,clip,width=8.5cm]{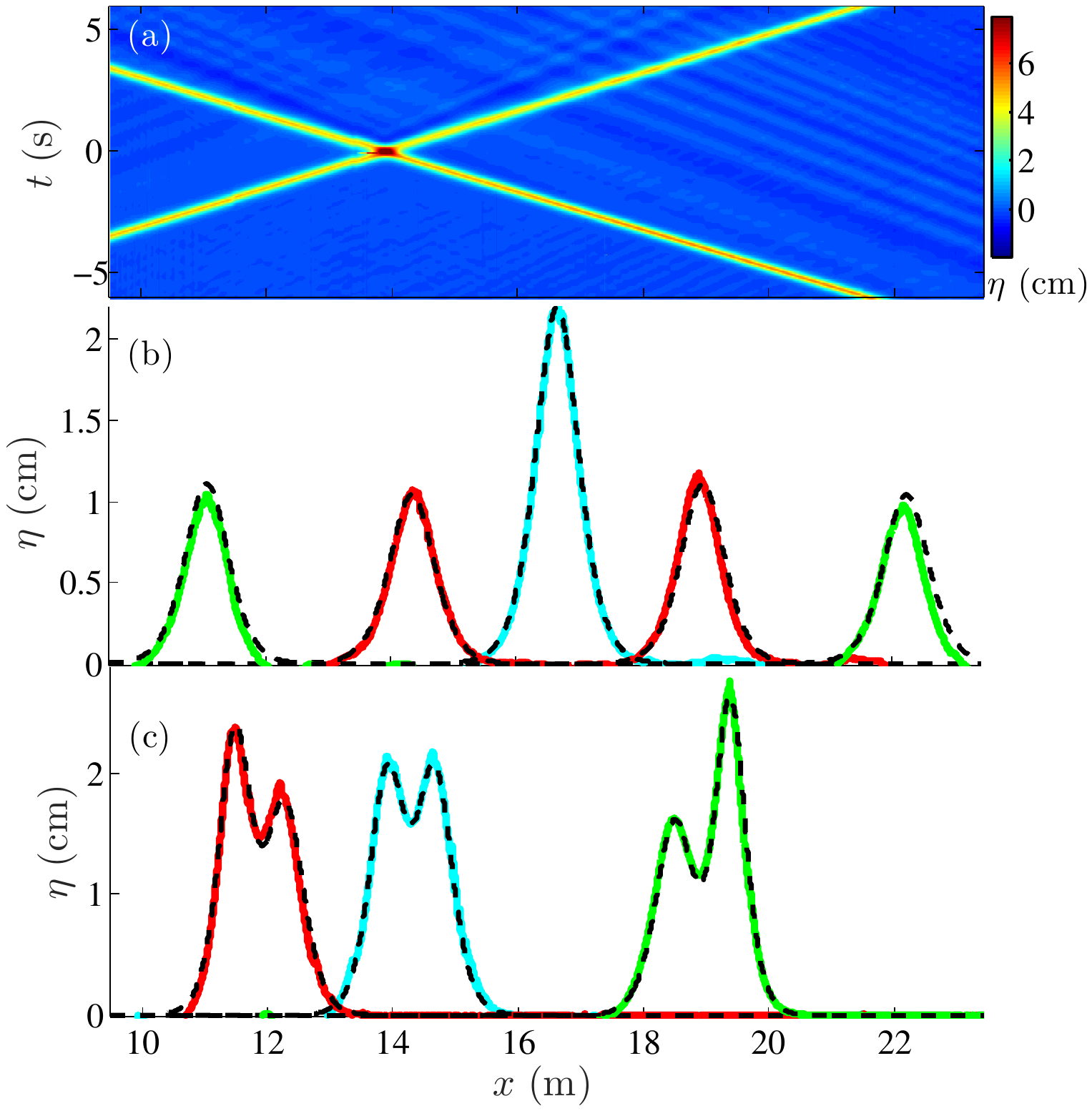}
\caption{Two-soliton interaction from experiment: (a) head-on  interaction of large amplitude solitons 
in a space-time representation. 
The wavemaker is located at $x=0$~m. Small dispersive waves radiated from the collision can be seen in the background, a sign of small departure from integrability. 
(b) Snapshots of head-on interaction of 
small amplitude solitons ($a/h=0.1$) traveling in opposite directions. 
Red: 2~s before the collision, blue: at the collision, green: 5~s after. 
Black dashed lines: solution of the KB equations (see text for details). 
(c) Collision of solitons traveling in the same direction (overtaking interaction), same colour code as in (b).}
\label{coll}
\end{figure}

Before discussing the realization of the soliton gas, we analyze the two-soliton solution in the laboratory.
{A space-time representation of a head-on collision of large amplitude solitons ($a/h \simeq 0.4$, with $a$ the amplitude of each soliton) is shown in Fig.~\ref{coll}(a). 
The left-going soliton is generated first and reflects at the right end wall.
}
The collision appears quasi-elastic and the resolution of the picture does not allow us to visualize here the phase shift induced by the collision.
The value $a/h$=0.4 is below the critical threshold, $(a/h)_c$= 0.6, for which the head-on collision leads to the formation of a residual jet, see \cite{chambarel2009head}.
Very small amplitude dispersive waves that move slower than the solitons  are emitted after the collision, see \cite{Craig,Chen:2014ea,Su:1980vb}. 
These waves, that are a sign of nonintegrability, remain of small amplitude and are only visible for collisions of very large solitons. 
Fig.~\ref{coll}(b) shows the head-on collision and Fig.~\ref{coll}(c) the overtaking collision for small amplitude solitons together 
with the two-soliton solutions of the KB eqs (\ref{KB}). The only free parameters are the amplitudes of the solitons that are obtained by extracting 
the amplitude of each soliton when they are widely separated (before and after the collision) and extrapolated to the collision point by compensating the dissipation. 
The two-soliton theoretical solution  matches nicely both experiments. Again, the observed short term dynamics is well reproduced by the integrable model. 

In order to obtain a bidirectional soliton gas, a large number of solitons must be injected in the wave flume. We face a technical difficulty 
when trying to inject individual solitons: the piston must recede slowly so as to not induce undesired waves and then move forward quickly to generate the soliton. 
The receding phase is so long that only a few solitons can be introduced in the entire flume by this procedure. 
{We circumvent this issue by continuously forcing a sinusoidal wave at the wave maker}. 
As observed experimentally in~\cite{Zabusky:2006en,Trillo} and numerically in~\cite{Salupere:2002uv,*salupere2003long,Kurkina:2016fb}, 
a sine wave in shallow water spontaneously steepens and decomposes,  after a certain propagation distance, into a train of solitons of various amplitudes. 
Our generation setup is similar to \cite{Ezersky,Ezersky2} but with a longer flume to ensure soliton fission. 
These solitons then interact with solitons emitted earlier that survive until dissipated by viscosity. 
The number of solitons depends proportionally on the Ursell number $U=\frac{3A\lambda^2}{16\pi^2h^3}$  
that measures the dispersive effects over the nonlinear ones ($A$ is taken as twice the standard deviation of $\eta$ and $\lambda$ is the wavelength), 
see~\cite{Trillo,Osborne} for details. In this way, we can inject a large number of solitons in the flume and their amplitude 
and density can be tuned by changing the amplitude and frequency of the sinusoidal forcing. 

\begin{figure}
\includegraphics[viewport=55 150 535 705,clip,width=8.5cm]{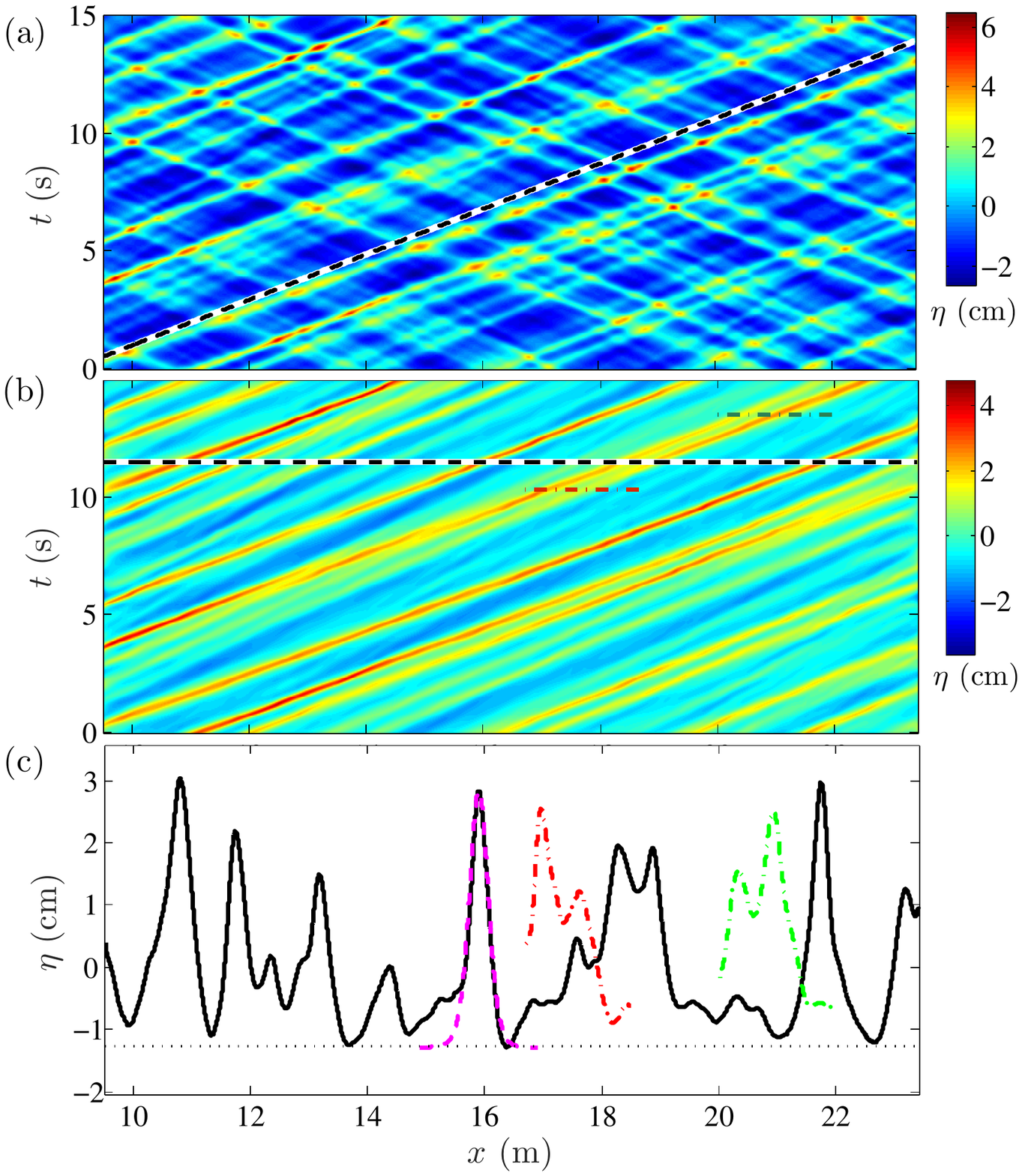}
\caption{(a) Space-time representation of the soliton gas. The horizontal scale is the same for all subfigures. Solitons traveling in both directions are visible. 
Many head-on collisions can be seen. The dashed line is the {long wave phase velocity $c_0=\sqrt{gh_{\!R}}$,} with $h_{\!R}$ the reference level of soliton propagation. 
(b) Part of the signal propagating away from the wave maker (to the right) obtained by the Hilbert transform. 
(c) Extraction of the surface elevation profile corresponding to the dashed line in (b) at $t=11.5$~s. 
Magenta: fit of a single soliton solution to an isolated pulse. The horizontal dotted line is the corresponding reference level $h_{\!R}$. 
Red: the surface elevation at $t=10.3$~s corresponding to dashed line in (b). Green:  surface elevation at $t=13.5$~s corresponding to the short dashed line in (b). 
These two curves correspond to pre- and post-collision states of a collision that occurs at $t=11.5$~s and $x=19$~m. 
They illustrate an example of overtaking soliton interaction in the soliton gas. 
}
\label{gaspar}
\end{figure}

An example of wave elevation of the obtained soliton gas can be seen in Fig.~\ref{gaspar}(a) in a space-time representation. The density of solitons is large enough so that solitons often interact. One can clearly see solitons that propagated in both directions as bright straight ridges, either increasing or decreasing in the $(x,t)$ space. The lines are not parallel due to the distribution of amplitudes of the solitons and because larger solitons propagate faster than smaller ones. The dashed line shows a line of slope equal to the shallow water linear long wave phase velocity $c_0=\sqrt{gh_{\!R}}$ 
(where $h_{\!R}$ is the reference level for solitons, slightly below $h$, see e.g. \cite{osborne1986solitons,giovanangeli} and Fig.~\ref{gaspar}(c)). 
The nearby soliton clearly propagates faster than a linear long wave. The head-on collisions are visible by the fact that the amplitude is maximum at each crossing of two counter-propagating solitons (as in Fig.~\ref{coll}(b)). Note that, although the solitons are injected periodically, there is no obvious sign of such periodicity in the plots. The large number of interactions among solitons seems to randomize the gas.

Waves propagating towards positive  $x$ can be separated from those propagating towards negative $x$ by 
{computing
the time-space Fourier transform $\tilde{\eta}(k,\omega)$ of the measured wave field $\eta(x,t)$}. 
Waves with $k>0$ \& $\omega>0$ (or $k<0$ \& $\omega<0$ as the field is real) 
propagate towards negative $x$ and waves with $k>0$ \& $\omega<0$  to positive $x$. 
Fig.~\ref{gaspar}(b) shows the time-space representation of the waves traveling with increasing $x$. 
Indeed, only the solitons going to the right are retained. An example of overtaking interaction of a large soliton 
and a smaller one is highlighted in Fig.~\ref{gaspar}(c) which strongly resembles the isolated collision shown in Fig.~\ref{coll}(c). 
The observation of such events supports the fact that our wave field is indeed dominated by solitons.

\begin{figure}
\includegraphics[viewport=175 200 510 590,clip,width=8.5cm]{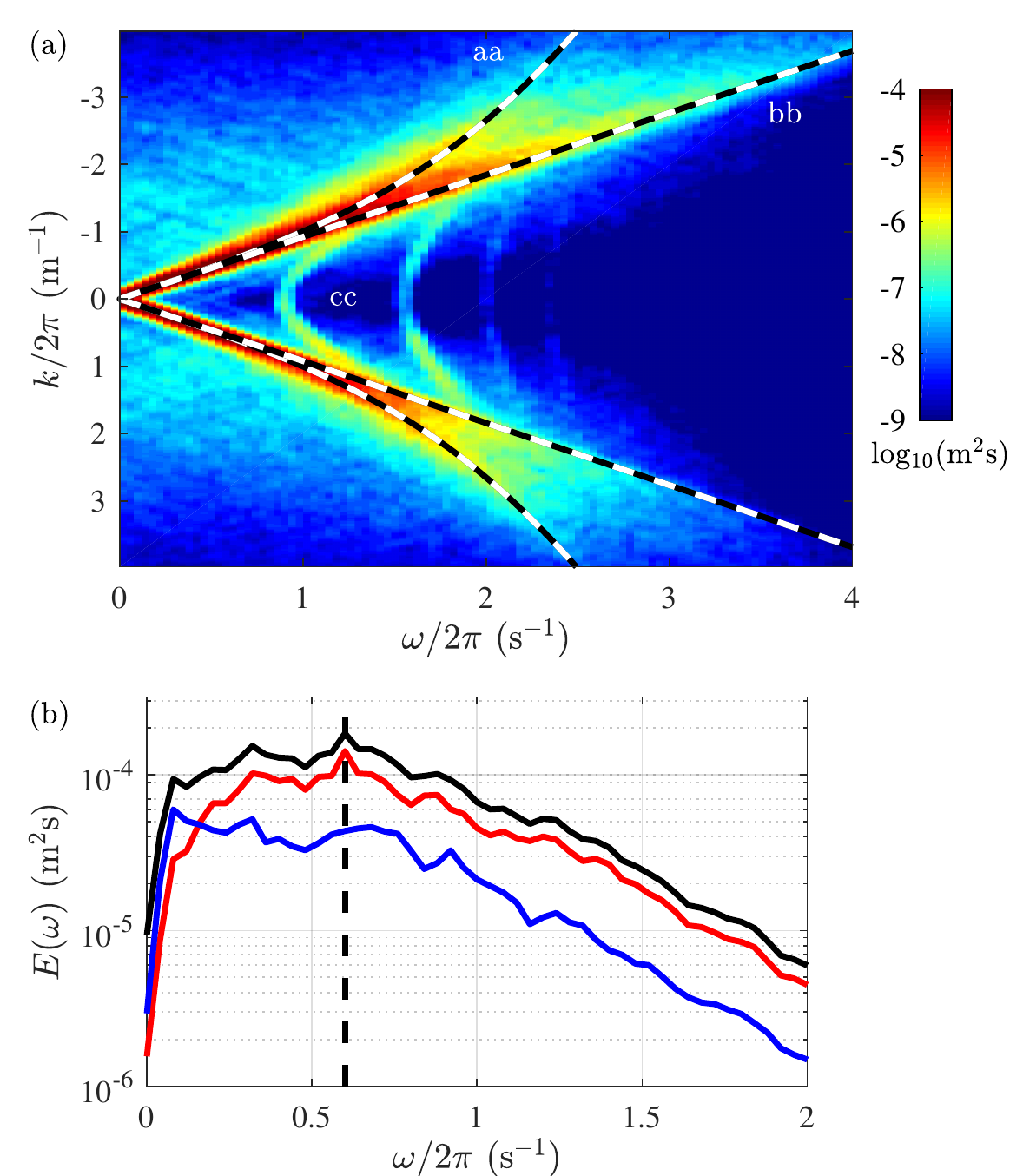}
\caption{(a) $E(k,\omega)$ spectrum of the soliton gas. Dashed curves: `aa' is the dispersion relation of linear waves $\omega=\sqrt{gk\tanh(kh)}$. `bb' is the relation $\omega=c_0 k$. `cc' shows the part of the spectrum related to transverse waves. (b) Frequency spectrum of the wave elevation average spatially over the locations accessible to the cameras. Black line: full signal. Red and blue lines: signal propagating to the right and to the left, respectively. The vertical dashed line corresponds to the forcing frequency at $0.6$~Hz.
}
\label{spec}
\end{figure}

A way to investigate all possible excitations in the system is to compute the space-time Fourier spectrum, $E(k,\omega)$, of the wave elevation. 
Here the Fourier transform is taken over the entire available spatial window (14~m long) and 
using the Welch method in time (i.e. using successive temporal windows of duration 25~s). 
The spectrum is shown in Fig.~\ref{spec}(a): several curves with a significant level of energy can be identified. 
The strongest is a symmetric pair of straight lines (labeled `bb'). They correspond to solitons whose Fourier components travel all at almost the same phase velocity. 
The thickness of the lines is due to the fact that solitons of various size coexist and travel at slightly distinct velocities depending on their amplitude. 
A second set of curves is labeled as `aa'. They correspond to linear dispersive waves following closely the dispersion relation curve $\omega=\sqrt{gk\tanh(kh)}$. 
As mentioned above, these waves may be radiated during collisions and are a signature of non integrability; also dissipated solitons may end up as linear dispersive. 
The lines marked `cc' correspond to very small standing transverse modes that are hard to avoid in wave flumes. 
The most energetic lines visible are a clear sign of a soliton gas; to better visualize the spread of energy across the soliton line, 
we compute the frequency spectrum $E(\omega)$  averaged spatially
, {see Fig.~\ref{spec}(b)}. The spectrum is shown for the full field (black line), 
the waves propagating towards positive $x$ (red line) and those reflected (blue line). In the curve corresponding to the waves leaving the wavemaker, 
a small peak at the forcing frequency is present. Energy is also present at frequencies lower and higher than this forcing. 
The high frequency content appears to decay exponentially; 
this result is consistent with numerical simulations of pure integrable equation containing a large content of solitons, see {\cite{Pelinovsky}}. 
The spectral band below the forcing frequency is rather flat; this is especially true for the case of {left-going waves with no forcing peak visible}. 
Amongst the solitons moving away from the wavemaker (right-going) are ``old'' solitons reflected on the piston and new ones emerging from the sinusoidal wave fission. 
The latter are responsible for the forcing peak as they have not experienced much collisions and have kept  the memory of their initial phase coherence. 
{In contrast the blue curve relates to solitons having undergone at least one reflection and traveled a longer distance, thus having experienced more interactions}.
These interactions have randomized their phase. Our observation of a flat spectrum for low frequencies is consistent with numerical simulations 
of the KdV equations (unidirectional soliton gas) described in \cite{Pelinovsky}. 
Field data measured at Duck Pier, North Carolina, are also consistent with a flat spectrum in a regime which is interpreted as dominated by solitons~\cite{Osborne}.  
A flat spectrum is actually reminiscent of the shot noise {due to discrete carriers in electronics or photonics. The electrons or photons} have a random distribution 
in time that leads to a flat spectrum~\cite{Papoulis}. All these results are in contrast with the observations of Costa {\it et al.}~\cite{Costa} 
who showed that a dense soliton gas in a narrow and shallow sound has low frequency power spectra that behave as $\sim \omega^{-1}$. 
%
A deeper analysis of our data would require to perform the Inverse Scattering Transform (IST)~\cite{Osborne,Smirnov} for estimating the content of the signal in terms of solitons or cnoidal waves. 
It is expected that such an approach might help to understand 
the low frequency power spectra trend in our experiments.

Further statistical information can be obtained from the skewness
 $S={\langle(\eta-\langle\eta\rangle)^3\rangle}/{\langle (\eta-\langle\eta\rangle)^2\rangle^{3/2}}$ and kurtosis $K={\langle(\eta-\langle\eta\rangle)^4\rangle}/{\langle (\eta-\langle\eta\rangle)^2\rangle^{2}}$ 
of the surface  elevation of waves travelling to positive $x$; those moments of the probability density function of the surface elevation
measure the departure from Gaussian statistics ($S=0$, $K=3$).
For the dataset shown in Fig.~\ref{spec},  $U=0.93$, in the case of waves propagating towards positive $x$, we obtain $S=0.9$, $K=3.45$; 
those numbers are very close to the numerical values  $S=0.8$, $K=3.45$ obtained  at $U=0.95$ in~\cite{Pelinovsky}. 
Although our setup is not integrable, many features appear consistent with numerical simulations of the integrable KdV equation. 
This is most likely due to the fact that dissipation operates over very long times scales (90~s, see Supplemental Material) compared to collision timescales 
(about 4~s from Fig.~\ref{gaspar}(b)). These collisions are the elementary process that leads to the randomization of the relative phases of the solitons. 
The observed largest soliton is approximately 5 cm high and  the progressive damping ensures that all soliton amplitudes, from 5 cm down to zero, are present in the flume. 
This  is characteristic of a warm soliton gas~\cite{El} and promotes many strong interactions as the one highlighted in Fig.~\ref{gaspar}(c).
If sufficient scale separation is present between collision and dissipation time scales, then the dynamics is ruled by the integrable wave equation. 

An open question is the role of head-on collisions on the statistical properties of the surface elevation. 
For each head-on collision, the resulting maximum amplitude exceeds that calculated by the superposition of the incident solitary waves \cite{Chen:2014ea}. 
Thus one may expect that the extreme events of the bidirectional gas to be of larger magnitude than that of the unidirectional KdV gas. 
Such analysis is left for future investigations.

In conclusion, in the present Letter we have given evidence of a hydrodynamic soliton gas 
 produced in the lab. Despite the fact that dissipation unavoidably exists in the flume, we were 
 able to produce a gas in a steady state regime whose properties are consistent with the conservative integrable dynamics. It is remarkable that the energy injected exits the wave system without changing the global  picture of integrable turbulence. 
We hope that our results will stimulate new theoretical and experimental work in other nonlinear dispersive media, such as optics, BEC etc, where soliton dynamics plays an important role.
\\

This project has received funding from the European Research Council (ERC) under the European Union's Horizon 2020 research and innovation programme (grant agreement No 647018-WATU).
M. O. has been funded by Progetto di Ricerca d'Ateneo CSTO160004. M.O. was supported by the ``Departments of Excellence 2018-2022'' Grant awarded by the Italian Ministry of Education, University and Research (MIUR) (L.232/2016). P. Suret, S. Randoux and G. El are acknowledged for discussions during the early stage of the work.
\bibliography{biblio}

\end{document}